\documentstyle[aas2pp4]{article}
\newcommand{\eg}{e.$.\!$g.\ } 
\newcommand{\be}{\begin{equation}}
\newcommand{\ee}{\end{equation}}
\newcommand{\etal}{ et al.\ }
\newcommand{\ie}{i$.\!$e$.\!$, }

\newcommand{\unit}[1]{\,{\rm #1}}
\newcommand{\hubble}[1]{H_0={#1}\unit{km\,sec^{-1}\,Mpc^{-1}}}

\newcommand{\qz}{q_0}
\newcommand{\labequn}[1]{\label{eq:#1}}
\newcommand{\labtab}[1]{\label{tab:#1}}
\newcommand{\labfig}[1]{\label{fig:#1}}
\newcommand{\labsecn}[1]{\label{sec:#1}}

\newcommand{\mean}[1]{\left\langle#1\right\rangle}
\newcommand{\equn}[1]{Equation~\ref{eq:#1}}
\newcommand{\tab}[1]{Table~\ref{tab:#1}}
\newcommand{\fig}[1]{Figure~\ref{fig:#1}}
\newcommand{\secn}[1]{Section~\ref{sec:#1}}

\begin{document}
\title{Correlations among  global photometric properties of disk galaxies}
\author{Habib G. Khosroshahi \altaffilmark{1}\affil{Institute for Advanced Studies in Basic Sciences,
P. O. Box 45195-159 Zanjan, Iran}   Yogesh Wadadekar
\altaffilmark{2} and Ajit Kembhavi \altaffilmark{3}
\affil{Inter University Centre for Astronomy and Astrophysics, Post Bag
4, Ganeshkhind, Pune 411007, India}}
\altaffiltext{1}{khosro@iasbs.ac.ir}
\altaffiltext{2}{yogesh@iucaa.ernet.in}
\altaffiltext{3}{akk@iucaa.ernet.in}
\begin{abstract}

Using a two-dimensional galaxy image decomposition technique, we
extract global bulge and disk parameters for a complete sample of
early type disk galaxies in the near infrared K band. We find
significant correlation of the bulge parameter $n$ with the central
bulge surface brightness $\mu_b(0)$ and with effective radius $r_e$. Using bivariate analysis techniques, we find that $\log n$,
$\log r_e$ and $\mu_b(0)$ are distributed in a plane with small scatter.
We do not find a strong correlation of $n$ with bulge-to-disk
luminosity ratio, contrary to earlier reports. $r_e$ and the disk
scale length $r_d$ are well correlated for these early type disk
galaxies, but with large scatter. We examine the implications of our
results to various bulge formation scenarios in disk galaxies.
\end{abstract}

\section{Introduction}

Accurate photometric and kinematic modeling of galactic morphology is
a prerequisite for tackling the many unresolved problems regarding the
formation and evolution of galaxies, \eg the formation of bulges in
disk galaxies. There are two popular scenarios for the formation of
such bulges: In the first scenario, the bulge and the disk are formed
independently, in the second the disk forms first and the bulge forms
later by secular evolution.  A quantitative evaluation of the
plausibility of such formation scenarios requires accurate extraction
of global parameters that describe the luminosity and density
distribution of the bulge as well as the disk.

Following the path breaking work of de Vaucouleurs (1948), the
empirical law named after him was extensively used to describe
the light distribution in elliptical galaxies and in the bulges of
spiral galaxies (de Vaucouleurs 1959).  With the advent of more
accurate CCD surface photometric techniques, systematic deviations
from the $r^{1/4}$ law for ellipticals were extensively reported in
the literature (Michard 1985, Schombert 1986, van den Bergh 1989,
Bingelli \& Cameron 1991). Alternative descriptions that emerged
included the $r^{1/n}$ law, with $n$ as a free parameter, first
proposed by Sersic (1968). This generalized de Vaucouleurs law was
used to fit a sample of elliptical and S0 galaxies by Caon, Cappacioli
\& D'Onofrio (1993) who found a large range of values in $n$, with $n$
correlated with the effective radius and the total luminosity of the
galaxy. More recently, Andredakis, Peletier \& Balcells (1995,
hereafter APB95) used the $r^{1/n}$ law to fit the bulges of spiral
galaxies. They found that $n$ varies systematically from values around
1 (which corresponds to an exponential) to 6, from late to early type
bulges; the profiles fall off less steeply near the center (i.e. have lower
$n$ values), in the later types. de Jong (1996) also reported that the
bulges of late type galaxies are better fit by an exponential than the
de Vaucouleurs' law.

The standard description for the light profile of the disk component
is an exponential, first proposed by Freeman (1970). It has been found
that an exponential profile also describes the light distribution in
the disks of S0s (Burstein 1979). An inner truncated exponential disk
has been proposed (Kormendy 1977a) to describe the disk profile, but is
not widely used.

After the photometric parameters describing the bulge and disk have
been extracted, the next step is to look for correlations between
them. Such correlations have been investigated by several authors (\eg
Kormendy 1977a; Schombert \& Bothun 1987; APB95; Courteau, de Jong \&
Broeils 1996; de Jong 1996; Bingelli \& Jerjen 1998). In this paper, we examine correlations
among the bulge and disk parameters for a complete magnitude limited
sample of high inclination disk galaxies using the $r^{1/n}$ law to
model the bulge and an exponential law to model the disk.

This paper is organized as follows: in \secn{data} we describe the
data set and the decomposition procedure used to obtain the bulge and
disk parameters. \secn{results} contains a detailed description of the
fitting and a comparison with values reported in the
literature. Various correlations observed between bulge and disk
parameters are discussed in \secn{cor}.
\secn{conclusions} contains the conclusions.

\section{The data}
\labsecn{data}
\subsection{Sample definition}

The galaxies that we use are from a complete, diameter-limited,
optically selected sample of early to intermediate type spirals from
the UGC catalogue. The sample was constructed by Balcells and Peletier
(1994), where details about selection criteria may be found. The sample is complete
for the specified galaxy types and diameter and magnitude
ranges. Thirty objects from this sample were studied by APB95; the
details of the observations are given in Peletier and Balcells (1997)
and APB95. We have used their reduced $K$ band images publicly available at the 
New Astronomy journal website\footnote{http://www.elsevier.nl/}, in this
work. The data at the website are identical to that described in APB95
except that they have been rebinned to a pixel scale of 0.549
arcsecond, which was the scale used for observations in the $UBRI$ bands.
The data reduction procedures are described in detail in Peletier \&
Balcells (1997) and APB95.

The main morphological criteria for the sample selection were that the
major axis diameter should be $> 2$ arcmin, and the axial ratio
in the B filter should be $> 1.56$, which corresponds to disk
inclinations $> 50$ degrees. Such a constraint on axial ratio
was necessary to the specific bulge disk decomposition technique
employed in APB95, which requires that the apparent ellipticities of
the bulge and the disk be as distinct as possible. The sample is thus
biased in orientation, towards edge-on galaxies. Of the 30 galaxies we 
have used only 26, for reasons cited in \secn{results}. The
morphological type index of these galaxies was obtained from the RC3
catalog (de Vaucouleurs \etal 1991) while the apparent magnitude $m$ and redshift $z$  were obtained from the NASA Extragalactic Database 
(NED). These quantities are listed in \tab{fitvalues}.

\subsection{The decomposition procedure}

Extracting the structural parameters of a galaxy requires the
separation of the observed light distribution into bulge and disk
components. There is considerable variation in the details of the
decomposition techniques proposed by various researchers. In recent
years, novel methods that employ two dimensional fits to broad band
galaxy images have been proposed (eg. Byun \& Freeman 1995; de Jong
1996; Wadadekar, Robbason and Kembhavi 1999) as
alternatives to conventional one dimensional techniques, where the
bulge and the disk models are fit either iteratively or simultaneously
to a one dimensional surface brightness profile (\eg Kormendy 1977a;
Kent 1985; Simien \& de Vaucouleurs 1986; Schombert \& Bothun 1987;
Baggett, Baggett \& Anderson 1998). Most of these decomposition
techniques assume specific surface brightness distributions like the de
Vaucouleurs' law for the bulge and an exponential distribution for the
disk. A notable exception is the method proposed by Kent (1985), which
can perform the decomposition in a model independent way. However,
extraction of global parameters that quantitatively describe the bulge
and the disk, still requires use of a model. APB95 use a two
dimensional generalization of Kent's method to extract their
parameters.

Our decomposition procedure is a full two dimensional method that uses
information from all pixels in the image. It essentially involves a
numerical solution to a signal-to-noise weighted $\chi^2$ minimization
problem. We achieve this minimization using the
Davidon-Fletcher-Powell variable metric algorithm included as part of
MINUIT -- a multidimensional minimization package from CERN (James
1994). Our technique involves building two dimensional image models
that best fit the observed galaxy images with the quality of the fit
evaluated by the $\chi^2$ value. The weights for the $\chi^2$ function
are computed using the $S/N$ ratio at each pixel of the galaxy
image. The model image is convolved with the measured PSF from the
galaxy frame before the $\chi^2$ is computed. Details of the accuracy
and reliability of the decomposition procedure, and the associated
galaxy simulation code, are provided in Wadadekar \etal (1999) and will not repeated
here.

Working with $K$ band data for decomposition is easier and more
effective than the visible bands, because of a relative lack of
absorption related features in the near infrared. The smooth,
featureless light profiles of galaxies in this band are very
convenient for extraction of global disk and bulge parameters.  Use of
$K$ band data is especially important for our sample, as it includes
several dusty galaxies. In our bulge-disk decomposition, we have the
following as free parameters: (1)~$I_b(0)$: the central bulge intensity,
in counts, which can later be converted to  $\rm mag\, arcsec^{-2}$
(2)~$r_e$: the half light radius of the bulge in pixels (3)~$e_b$: the
ellipticity of the bulge (4)~$n$: the bulge shape parameter (5)~$I_d(0)$: the
central intensity of the disk  in counts (6)~$r_d$: the scale length
of the disk in pixels (7)~$e_d$: the ellipticity of the disk

With these definitions the bulge intensity distribution can be written as  
\begin{eqnarray}
I_{bulge}(x,y) &=& I_b(0) e^{ -2.303 b_n (r_{bulge}/r_e)^{1/n}}, \\
r_{bulge} &=& \sqrt{x^2 + y^2/(1 - e_b)^2}, \nonumber
\end{eqnarray}  
where $x$ and $y$ are the distances from the center of the galaxy along the major 
and minor axis respectively and $b_n$ is a function of $n$ which is a  root of an equation
involving the incomplete gamma function.
However, a simplification can be introduced in the procedure because
$b_n$ can be expressed as a linear function of $n$, accurate to better 
than one part in
$10^{5}$, by
\begin{eqnarray}
b_n  &=& 0.868242\, n - 0.142058. \nonumber
\end{eqnarray} 
For $n=4$, which corresponds to de Vaucouleurs law, $b_4 = 3.33$.

The projected disk profile is represented by an exponential distribution,
\begin{eqnarray}
I_{disk}(x,y) &=& I_d(0) e^{- r_{disk}/r_d},\\
r_{disk} &=& \sqrt{x^2 + y^2/(1 - e_d)^2}. \nonumber
\end{eqnarray}
The ellipticity of the disk in the image is due to projection effects alone 
and is given  by
\begin{eqnarray}
e_d &=& 1 - \cos(i),
\end{eqnarray}
where $i$ is the angle of inclination between the line of sight and
the normal to the disk plane.

The bulge-to-disk luminosity ratio is a parameter which is commonly used as a
quantitative measure for morphological classification of galaxies. For
the $1/n$ law it is given by
\begin{equation}
(B/D)_n=\frac{n\Gamma (2n)}{(2.303b_n)^{2n}}\left( \frac{I_b(0)}{I_d(0)}\right) 
\left( \frac{r_e}{r_d}\right)^2.
\labequn{dbybn}
\end{equation} 
For $n=4$ this reduces to the familiar result from Mihalas and Binney (1981),
\begin{equation}
(B/D)_4= \frac{1}{0.28} \left( \frac{I_b(r_e)}{I_d(0)}\right) \left( \frac{r_e}{r_d}\right)^2,
\labequn{dbyb}
\end{equation}
Note that we use $I_b(r_e)/I_d(0)$ in the above
equation,  where $I_b(r_e) =10^{-3.33}\, I_b(0)$ is the intensity at the
half--light radius, in place  of $I_b(0)/I_d(0)$ used in \equn{dbybn}.

\section{Decomposition results}
\labsecn{results}

We were able to obtain satisfactory fits with a reduced $\chi_{\nu}^{\,2} < 2$
for 26 of the 30 galaxies in the APB95 sample. We were unable to get a
good decomposition for 4 galaxies: NGC 5389, NGC 5443, NGC 5577 and
NGC 5746. This is very likely because of the fact that all these
galaxies show significant departure from twofold symmetry, in the high
$S/N$ regions near the center of the galaxy. The constant (\ie
unchanging with radius) and equal position angle that we assume for
the bulge and disk components, which implies a twofold symmetry, is
not a valid assumption for these objects. Since our decomposition does
not give us a fit with a satisfactory reduced $\chi_{\nu}^{\,2}$ for
these galaxies we have not used them in our analysis. Our
subsequent discussion only applies to the remaining 26 galaxies, for
which we have been able to obtain a satisfactory fit. Our extracted
parameter values for these 26 galaxies are presented in
\tab{fitvalues}. Error on the bulge parameter $n$ represented by $\Delta n$
is also listed. This error was computed by MINUIT from the Hessian
error matrix, which is the inverse of the matrix of second derivatives of the
$\chi^2$ function, with respect to the variable parameters. 

Accurate estimation of the full width at half maximum (FWHM) of the
point spread function (PSF) is crucial to our decomposition procedure
because we give maximum weightage to the high $S/N$ points near the
centre of the galaxy. This is especially true if the scale length in
angular terms is comparable to the seeing FWHM. The central galaxy
pixels are affected the most by convolution with the (possibly
inaccurate) PSF.  The FWHM is usually computed using the brightest
unsaturated stars in the image.  In our data set only 12 galaxy
images contain suitable bright stars in the field and their FWHM
ranges from 0.9 to 1.2 arcsec. For the other 14 objects, it is not
possible to directly measure the FWHM. In order to estimate the FWHM
for these galaxies, we took test runs with the value of the FWHM set
to 0.8, 1.0 and 1.2 arcsecond and tried to fit all 14 galaxies with
these values to see how various parameters, especially those of the
bulge, are affected by an erroneous estimation of the PSF.  Changing
the FWHM within this range does not change the results
significantly. In the extreme case, changing from 0.8 to 1.2
arcsecond, the extracted parameters do not show a discrepancy larger
than about 10 percent.  This fortunate circumstance in our sample is
due to the fact that bulge effective radii are $\geq 3\times$ FWHM for
all except four galaxies which have bulge effective radii of about
1.75 FWHM. We used an FWHM of 1.0 arcsecond in our analysis of all 26
galaxies.

\subsection{Comparison with APB95 results}
We have compared parameter values from APB95 with those obtained in this
work for the same sample.  We use  $\hubble{50}$
and $\qz=0.5$ , throughout this paper. 
\fig{andourcomp} shows plots of values extracted by us compared to the
values reported in APB95 for each parameter. The diagonal line in
each plot represents a perfect match between the two sets of values.
   
Ellipticity of the bulge and disk are rather stable parameters in the
fitting process. These parameters are never affected by the choice of
initial values and can usually be extracted with a 
small error.  APB95 claim that it is not easy to extract bulge
ellipticities by any method, and errors of the order of 30 percent
are possible. As we have already pointed out in Wadadekar \etal (1999), our
decomposition method allows us to extract bulge ellipticities with
much better accuracies (to better that 5 percent accuracy in 95
percent of simulated test galaxies). We wish to emphasize here  that
our technique is different from that used in APB95 and
it does not require the precondition that apparent ellipticities of
bulge and disk be as distinct as possible.

Comparison of extracted values of the shape parameter $n$ of the bulge
shows that for 14 galaxies out of 26, our error bar for $n$ overlaps with
the error bar on the corresponding $n$ values reported in APB95. The
deviation for most of the remaining galaxies is not large.

The bulge effective radius is consistently estimated overall. However, for
the larger bulges, our value is systematically higher than the APB95
value by about 20 percent.

The authors of APB95 have not reported the values of the disk scale
length in their paper. However, in a subsequent paper presenting their data
(Peletier \& Balcells 1997), they employed a different technique to
measure the disk scale length. They extracted the disk scale length
from the disk profile computed by azimuthally averaging the total
light of the galaxy (including light from both the bulge and the disk 
components), in
a wedge with full width of 10 degrees, placed slightly away from each
semi-major axis. Such a technique leads to contamination from
the bulge light in the central parts. This ``extra'' light at the center
leads to a consistent underestimation of disk scale length by their
technique, as shown in panel (d) of \fig{andourcomp}.
   
\section{Correlations} 
\labsecn{cor}

The next step in testing of predictions made by theoretical models for
the formation and evolution of galaxies, is to look for correlations
between global photometric parameters.  Several such correlations have
been reported in the literature. These correlations are useful in
reducing the number of free parameters that characterize galaxies,
making the analysis easier. One such correlation was found by Kormendy
(1977b) for large ellipticals by fitting the galaxy profiles with the
$r^{1/4}$ law. It relates the central surface
brightness $B_0$ in the $B$ band measured in $\rm mag\, arcsec^{-2}$
to the half light radius $r_e$, 
\be
B_0 = 3.02 \log r_e + 19.74\, .
\labequn{kormendy}
\ee
The  constants in this equation were determined by Kormendy using a least square
fit to the galaxy profiles. Subsequent work (eg. Djorgovski \& Davis 1987) found 
that the  mean surface brightness within the effective
radius,  $\mean{SB}_{eff}$, was a better parameter to use than the central
surface brightness in the Kormendy relation, as 
it is affected less by measurement errors. Such a relation is free of
model dependent parameters because both $\mean{SB}_{eff}$ and $r_e$ can be measured operationally,
without assuming a specific form for the luminosity profile of the
galaxy. 

For
a de Vaucouleurs elliptical, it can be shown that
\be
L \propto I_b(0) r_e\,^2 \propto I_b(r_e) r_e\,^2  ,
\labequn{luminositydev}
\ee
where $L$ is the luminosity and $I_b(r_e) = 10^{-3.33}\, I_b(0)$. For galaxies obeying \equn{kormendy} and
\equn{luminositydev}, 
the luminosity of elliptical galaxies  reduces to a
one parameter family. For the $1/n$ law the luminosity is
given by
\be
L \propto I_b(r_e) r_e\,^2\, \psi(n),
\labequn{luminosityn}
\ee
where 
\be
\psi(n) = \frac{n\Gamma (2n) e^{2.303b_n}}{(2.303b_n)^{2n}}. 
\ee
Using the Kormendy relation, the luminosity of elliptical
galaxies now reduces to a two parameter family.

In recent years, there have been attempts to correlate the $n$
parameter of bulges with other observables, with mixed success. Caon
\etal 1993 reported a correlation of $n$ with luminosity as well
as effective radius, while APB95 reported a correlation between $n$ and
morphological type and between $n$ and the $B/D$ luminosity ratio. The parameter
$n$ measures the steepness of the luminosity profile of the bulge.
This is illustrated in \fig{nprofiles}, where we present surface
brightness profiles in arbitrary units for a $r^{1/n}$ galaxy with
$I_b(0)=1$ for $n= 1,4,7$.  It is seen that as $n$ increases, the
steepness of the profile increases rapidly near the centre, while for
$ r > r_e$ the profiles become flatter with increasing $n$. If the
value of $n$ is very large, say $n=16$, the largest value reported by
Caon
\etal (1993) for giant ellipticals, the profile is so steep that a
very large value of $I_b(0)$ would be required to prevent the galaxy from
disappearing into the background within a very short distance from its
centre.  From \equn{luminosityn} it follows that a galaxy with high
$n$ would have a high luminosity such as is found only among giant
ellipticals. Such a trend has been observed in a sample of 119 Abell
brightest cluster galaxies by Graham \etal (1996).

We shall first consider correlations among bulge and disk parameters
separately and then correlations involving both bulge and disk parameters.

\subsection{Correlations among bulge parameters}
\subsubsection{Univariate correlations}

Here we consider the three bulge parameters $n$, $\mu_b(0)$ and $r_e$,
taken two at a time.  We find that there is good correlation between
effective radius and $n$ as shown in
\fig{nrecor}. The linear correlation coefficient is 0.61 with a
significance level of 99.98 percent.  There is an even stronger
correlation between $n$ and the bulge central surface brightness as
shown in \fig{ni0cor}. The linear correlation coefficient in this case
is -0.88 at a significance level better than 99.99 percent.  There is
no significant correlation between $\mu_b(0)$ and $r_e$.  The Kormendy
relation between $\mean{SB}_{eff}$ and $r_e$ would imply a correlation
between $\mu_b(0)$ and $r_e$, if a simple offset exists between
$\mu_b(0)$ and $\mu_b(r_e)$ as in ellipticals following the de
Vaucouleurs law.  When the parameter $n$ is introduced, the relation
between $\mu_b(0)$ and $\mu_b(r_e)$ is strongly dependent on $n$
because $\mu_b(0) = \mu_b(r_e) -2.5\, b_n$ and therefore there is no
correlation between $\mu_b(0)$ and $r_e$, in spite of the Kormendy
relation.

\subsubsection{Bivariate correlations and a new fundamental plane?}

Since univariate correlations exist but with large scatter, there is
the possibility that some of the scatter may be caused by the effects
of a third parameter. 

The methods of multivariate statistics can be applied here to the three
bulge parameters $n$, $\mu_b(0)$ and $r_e$. We obtain a least square fit of
two of these quantities and correlate residuals from the fit with the
third quantity. If the residuals show significant correlation with the
third quantity, we make a linear combination of that quantity and one
of the previous two, and optimize the fit again. We have used the
MINUIT minimization program (James 1994) for this optimization. The
quality of the fit is judged by the $\chi^2$ value and the value of
the linear regression coefficient. After such bivariate analysis using
all possible combinations for the three variables, we find that the
best fit is obtained by expressing $\log n$ as a linear combination of
the other two parameters. The best fit plane is:
\begin{eqnarray}
\log n &=& (0.130 \pm 0.040) \log r_e - (0.073 \pm 0.011)
\mu_b(0)\nonumber\\
 &+& (1.21 \pm 0.11)
\labequn{bicor}
\end{eqnarray}
which corresponds to a scaling law:
\begin{equation}
n \sim  I_b(0)^{0.183 \pm 0.028}\, r_e^{0.130 \pm 0.040}
\labequn{bivarscaling}
\end{equation}
The fit is shown in \fig{bicor}. The scatter in the plot is 
small, 0.058 dex in $\log n$. This is equivalent to a scatter of $2.5 
\times 0.058 = 0.145$ in magnitude units.

We explored to what  extent  these correlations are caused by a 
coupling of the errors associated with the model parameters due to the 
 parameters compensating each other, as each parameter changes. If
such is the case, the bivariate correlation would not be as significant as it
appears. We used a  Monte-Carlo like simulation technique, where we
simulated galaxy images with the same parameter distributions as that of the
sample  galaxies. We tested to  see to what extent a correlation
appears in the fitted parameters, when  there is no correlation in the
input parameters to the simulation program. We found that no
artificial univariate or bivariate correlations are generated
(Khosroshahi, Wadadekar \& Kembhavi, under preparation).

Such a tight correlation involving these three purely photometric
parameters has not been noticed previously and is the main result of
this paper.

\subsection{Univariate Correlations among disk parameters}

The light distribution in the disk is defined by just two parameters,
the disk central surface brightness, $\mu_d(0)$ and
$r_d$. An anti-correlation is seen between these two quantities as
shown in
\fig{isrscor} but the scatter is large. The linear correlation coefficient
between $\mu_d(0)$ and $\log r_d$ is 0.645 with a significance of 99.96
percent.

\subsection{Bulge-disk correlations}
\subsubsection{$\log n - \log (B/D)$ correlation}
A strong correlation of $\log n$ with $\log (B/D)$ with a correlation
coefficient of 0.54 and significance of 99.7 percent for 30 galaxies
has been reported in APB95.  In our work, we find that the $\log
n-\log(B/D)$ correlation is weak with a linear correlation coefficient
of 0.30 and a significance level of only 86.5 percent (See
\fig{nbdcor}). Note that these results are obtained when we consider
the 26 galaxies for which we have a good fit. Using APB95 results for
the same 26 galaxies, we get a correlation coefficient of 0.55 with a
significance level of 99.7 percent. We believe that the marked
difference between our results and those of APB95 may be due to the fact
that APB95 used a different technique to estimate the disk
contribution. The disk magnitude used in their $B/D$ luminosity ratio was
determined by straightforward integration of the disk profile until
the last measured point. The disk profile was obtained by simply
subtracting the bulge profile from the total major axis profile. In
such a technique any errors in determination of the bulge profile
would get transferred to the disk profile directly. 

A better correlation is seen between $\log n$ and bulge luminosity. 
 We show such a plot in
\fig{nabmagcor}, where the absolute $K$ magnitude of the bulge is used
as a measure of its luminosity. This correlation is tighter than our
$\log n-\log(B/D)$ correlation with linear correlation coefficient
-0.40 with a significance of 95.9 percent. However the scatter is
quite large.

We have shown in \fig{nmorcor} a plot of $\log n$ against the
morphological type taken from the RC3 (de Vaucouleurs \etal 1991)
catalog. Since our sample consists of edge-on galaxies, it is
difficult to determine their morphological type using criteria
involving spiral arms. In spite of this problem a trend is seen with the
later morphological types having smaller values of $n$. But the formal
correlation coefficient between $\log n$ and morphological type index $T$ is
low, essentially because of the cluster of points with high $n$ and
$T=3$. APB95 found a tighter correlation between the two parameters
(see their Figure 5(a)), but much of the correlation there is due to
the data points from Kent (1986), where the morphological
classification is more reliable because the galaxies are less
edge-on.

\subsubsection{$r_e - r_d$ correlation}
An interesting correlation between bulge and disk scale lengths was reported by Courteau, de Jong and
Broeils (1996). They used two
samples of late type spiral galaxies, one from Broeils and Courteau
(1996) and the other from de Jong (1996). They modeled
the bulge keeping the shape parameter fixed at $n=1$ and obtained the
bulge scale length $r_b$  for their exponential fit. For $n=1$, $r_b$ is related to the effective radius $r_e$ by $r_e = 1.679\, r_b$.
For 83 galaxies from the de
Jong sample they found a mean $r_b/r_d$ ratio in the $R$ band 
of $0.072 \pm 0.043 $
(equivalent to a mean  $r_e/r_d$ ratio of $0.121 \pm 0.072$) 
whereas for 243 galaxies from the Broeils and Courteau sample they
found a mean $r_b/r_d$ ratio of $0.082 \pm 0.053$ (equivalent to a
mean $r_e/r_d$ ratio of $0.138 \pm 0.089$), also in the $R$ band. Such a relationship
between bulge and disk scale lengths is best understood in models
where the disk forms first and the bulge emerges later from the
disk. In such models, the exponential disk is formed by the
redistribution of angular momentum by viscous evolution, provided that
star formation occurs on roughly the same timescale.  A bulge emerges
naturally in such a scheme and its properties depend only on the
relative timescales of star formation and viscous transport and on the
total angular momentum (Combes \etal 1990; Saio \& Yoshi 1990;
Struck-Marcell 1991). Correlated bulge and disk scale lengths are
predicted by these models (Courteau \etal 1996).

A comparison between our distribution of $r_e/r_d$ in the $K$ band
with earlier results can be made using results from de Jong (1996). He
obtained a mean value of $(r_e/r_d)_K$ of $0.15 \pm 0.05$, again
for $n=1$ fixed, for his sample of predominantly late type face-on disk
galaxies. The corresponding value for our sample of predominantly
early type edge-on disk galaxies is $0.33 \pm 0.17$. A Student's
t-test reveals that the means for the two samples are different a
significance level of 99.5 percent. This indicates that the ratio
$r_e/r_d$ is not constant over the entire population of disk
galaxies. A recent statistical analysis of the the de Jong sample by
Graham \& Prieto (1999) has demonstrated that the scale length ratio
does indeed vary with Hubble type.  These results lend observational
support to the hypothesis that more than one bulge formation
mechanism may be at work. As suggested by Courteau \etal (1996) it may 
be possible that early type and late type disk galaxies
were formed by different physical mechanisms - big bulges ($>2$ kpc as
in Sa galaxies) formed principally from a minor merger and smaller
bulges formed mainly via secular evolution.

In \fig{rersdejong}, we plot the bulge effective radius against the
disk scale length, with points from our sample shown as filled circles
and those from the $K$ band observations of de Jong (1996) as open
circles. From the plot it is apparent that the disk scale lengths of
our sample and the de Jong sample span the same range, but our bulge
effective radii are larger on the average (which is to be expected in
early type galaxies) giving us a mean $r_e/r_d$ ratio that is about
2.2 times larger than the de Jong value.

We next test the dependence of the scale length ratio on morphological
type. In \fig{rebyrdmorcor} we show a plot of the two
parameters. There is no indication in the plot of a dependence of the
scale length on morphological type and the formal correlation
coefficient is not significant. However, the scatter is large. Some of
this scatter could be due to incorrect morphological type
classification, due to the edge-on nature of the galaxies. The plot
should be reexamined with a sample of galaxies with more secure
morphological classification.

\subsubsection{Other Correlations}
A plot of $\log B/D$ 
against  morphological type in \fig{bdmorcor} shows an anti-correlation
which is significant at the 97.8 percent level. This is not surprising as
the variation of the $B/D$ luminosity ratio with morphological type is
supposed to be an important characteristic of the the Hubble sequence,
with the ratio decreasing towards the later Hubble types.

We have shown in \fig{i0byidmorcor} a plot of $\log(I_b(0)/I_d(0))$ against
morphological type for our sample. A clear dependence is seen which is 
particularly striking if we ignore the points at $T=3$. The linear
correlation coefficient for all 26 points is -0.36, which is significant at the 92.8
percent level. Omitting the eight points at $T=3$, the correlation
coefficient improves to -0.75 and the significance to 99.97
percent. The decrease in the $B/D$ luminosity ratio towards the later
morphological types is driven by this decrease in  $I_b(0)/I_d(0)$.
because $r_e/r_d$ does not depend strongly on morphological type as is 
seen in \fig{rebyrdmorcor}.

If the bulges of early and late type disk galaxies did indeed form by
different mechanisms, then there should exist other observable
differences in the bulge properties of early and late type disk
galaxies. We test this idea by plotting the surface brightness at the
effective radius $r_e$ as a function of bulge effective radius for our
sample which is dominated by {\em early} type disk galaxies and that
of de Jong (1996) which contains mostly {\em late} type disk
galaxies. \fig{sbredejong} shows our points as filled circles and the
points from de Jong (1996) as open circles. This plot can be viewed in
conjunction with Fig. 7(b) in APB95 to see a trend: bulges of early
type disk galaxies seem to obey a linear Kormendy type relation,
albeit with large scatter, suggesting a formation scenario similar to
that of ellipticals.  On the other hand, bulges of late type spirals
are widely scattered mostly below this line, suggesting a different formation
history, with secular evolution being one of the possibilities. We
will explore this trend further in a future paper using $K$ band data
on a sample of elliptical galaxies.

The authors of APB95 rejected secular evolution models on the 
basis of their continuous spectrum of the index $n$ versus
morphological type and $B/D$ luminosity ratio. They proposed that the smooth sequence they
observed, in their Fig. 5, is an indicator of a single mechanism of
bulge formation in all disk galaxies. We have found that neither $B/D$ 
luminosity ratio nor morphological type correlate with $n$ as strongly as APB95. Additionally, as pointed
out by Courteau \etal (1996), the scatter in the plot is too large to
justify the APB95 claim.  In such a situation, ruling out the
possibility of secular evolution is not warranted. The present data
are too sparse to test for any bimodality in the variation of $n$ with
morphological type.

Observationally these scenarios can be differentiated by studying the
evolution of $B/D$ luminosity ratio and colors of galaxies with redshift. A
series of schematic models have been developed using the data from
APB95 and de Jong (1996) by Bouwens, Cayon \& Silk
(1999). Unfortunately, currently available high $z$ galaxy
observations have small sample sizes and the measurement uncertainties
are too large to test the predictions of these models.
  
\section{Conclusions}
\labsecn{conclusions}

The bulge shape parameter $n$ is separately correlated with both the
bulge effective radius and the bulge central surface brightness. A
larger value of $n$ implies a larger bulge which is also brighter at the centre.  A
strong correlation between $n$ and the bulge-to-disk luminosity ratio
was reported in APB95. We do not find a correlation as strong as
theirs in our work.  A strong bivariate correlation between bulge
central surface brightness, bulge effective radius and $n$ exists with a
scatter as small as 0.058 dex in $\log n$. Such a tight best fit plane
is the first bivariate correlation noted involving these three purely
photometric parameters. The physical implications of this correlation
need to be explored by analytic and numerical techniques.  

The bulge
and disk scale lengths are correlated, but the scatter is large
indicating that more than one bulge formation mechanism may be at work
with a possible dependence on morphological type. In contrast to the
findings of Courteau \etal (1996) and in agreement with the
conclusions of Graham \& Prieto (1999), the mean scale length ratio
does seem to change with morphological type. The disk central
surface brightness and disk scale length are anti-correlated.

A detailed near infrared study of a complete sample of elliptical
galaxies is underway, as a complementary study to the present
work. The results from that study will enable us to compare and
contrast the photometric properties of the bulges of spirals and
ellipticals, and study in detail what effect the presence of a disk
has on the global parameters of the bulges of disk galaxies. It would
be interesting to see if a bivariate relation, of the form described
in this work, also holds for elliptical galaxies.

\acknowledgements
We thank Y. C. Andredakis, R.F. Peletier and M. Balcells for making 
their data publicly available. We thank S. George Djorgovski and the
referee, Alister Graham for several helpful comments and suggestions.

This research has made use of the NASA/IPAC Extragalactic Database
(NED) which is operated by the Jet Propulsion Laboratory, California
Institute of Technology, under contract with the National Aeronautics
and Space Administration.

\newpage

\centerline{REFERENCES}
\noindent Andredakis, Y.C., Peletier, R.F., \& Balcells, M. 1995,
\mnras, 275, 874\\
Balcells, M. \& Peletier, R.F. 1994, \aj, 107, 135\\
Baggett, W.E., Baggett, S.M., \&  Anderson, K.S.J. 1998, \aj, 116, 1626\\ 
Bingelli, B., \& Cameron, L.M. 1991, \aap, 252, 27 \\
Bingelli, B., \& Jerjen, H. 1998, \aap, 333, 17\\
Bouwens, R., Cayon, L. \& Silk, J. 1999, \apj, 516, 77\\
Burstein, D. 1979, \apjs, 41, 435 \\
Byun, Y.I., \& Freeman, K. 1995, \apj, 448, 563 \\
Caon, N., Capaccioli, M. \& D'Onofrio, M. 1993, \mnras, 163,1013\\ 
Combes, F., Debbash, F., Friedli, D., \& Pfenniger, D. 1990, \aap, 233, 82\\
Courteau, S., de Jong, R. S. \& Broeils, A. H. 1996, \apj, 457, L73\\
de Jong R. S. 1996, \aaps, 118, 557 \\
de Vaucouleurs, G. 1948, Ann. d'Astrophys., 11, 247 \\
de Vaucouleurs, G. 1959, Hdb. d. Physik, 53, 311 \\
de Vaucouleurs, G. {\em et al}, 1991, Third Reference Catalog of
Bright Galaxies (New York: Springer) (RC3) \\
Djorgovski, S.G. \& Davis, M. 1987, \apj, 313 , 59\\
Freeman, K. 1970, \apj, 160, 811\\
James, F. 1994, MINUIT: Function Minimization and Error Analysis (CERN 
Program Libr. Long Writeup D506) (version 94.1; Geneva: CERN)\\
Graham, A. W., \& Prieto, M. 1999, \apjl, 524, L23\\
Graham, A., Lauer, T. R., Colless, M. \& Postman, M. 1996, \apj, 465, 534\\
Kent, S.M. 1985, \apjs, 59, 115\\
Kent, S.M. 1986, \aj, 91, 1301\\
Kormendy, J. 1977a, \apj, 217, 406  \\
Kormendy, J. 1977b, \apj, 218, 333  \\
Michard, R. 1985, \aap, 59, 205 \\
Mihalas D. \& Binney J. 1981, Galactic Astronomy (New York:
W. H. Freeman and Company)\\
Peletier, R.F., \& Balcells, M. 1997, New Astronomy, 1, 349 \\
Saio, H., \& Yoshii, Y. 1990, \apj, 363, 40\\
Schombert, J. M. 1986, \apj, 60, 603 \\
Schombert, J., \& Bothun, G. D. 1987, \aj, 93, 60 \\
Sersic, J.L. 1968, Atlas de galaxias australes. Observatorio
Astronomica, Cordoba\\
Simien, F., \& de Vaucouleurs, G. 1986, \apj, 302, 564 \\
Struck-Marcell, C. 1991, \apj, 368, 348\\
van den Bergh, S. 1989, PASP, 101, 1072 \\
Wadadekar Y., Robbason R., \& Kembhavi, A. 1999, \aj, 117, 1219\\
\newpage

\begin{deluxetable}{l r l l l l l l  l l r r r}
\tablewidth{0pt}
\tablecaption{Extracted bulge and disk parameters}
\tablehead{\colhead{Name}&\colhead{$T$\tablenotemark{a}}&\colhead{$m$\tablenotemark{b}}&\colhead{$z$\tablenotemark{b}}&\colhead{$e_b$}&\colhead{$e_d$}&\colhead{$n$}& 
\colhead{$\Delta n$}&\colhead{$r_e$}&\colhead{$r_d$}&\colhead{$\mu_b(0)$}&\colhead{$\mu_d(0)$}&\colhead{$B/D$}\\&&&&&&&&(kpc)&(kpc)&&&} 
\startdata
IC 1029   & 3& 12.2  &0.00795&0.41 & 0.79 & 2.19 &   0.14  &  0.93 & 10.25 &10.96&16.94& 0.09 \\ 
NGC 5326  & 1& 12.9  &0.00841&0.31 & 0.54 & 1.66 &   0.10  &  0.63 &  2.19 &11.83&15.51& 0.30 \\ 
NGC 5362  & 3& 13.14 &0.00726&0.41 & 0.57 & 2.31 &   0.34  &  1.36 &  3.72 &12.92&17.04& 0.06 \\ 
NGC 5422  &-2& 12.81 &0.00594&0.23 & 0.77 & 2.41 &   0.07  &  0.71 &  3.47 &10.87&16.20& 0.18 \\ 
NGC 5475  & 0& 13.50 &0.00574&0.00 & 0.72 & 2.02 &   0.14  &  0.36 &  2.23 &11.69&15.96& 0.08 \\ 
NGC 5587  & 0& 13.51 &0.00768&0.22 & 0.78 & 2.11 &   0.15  &  0.67 &  5.93 &12.32&16.64& 0.04 \\ 
NGC 5675  & 0& 13.70 &0.01325&0.24 & 0.63 & 2.90 &   0.14  &  1.41 &  4.36 &10.37&16.28& 0.31 \\ 
NGC 5689  & 0& 12.8  &0.00720&0.47 & 0.54 & 3.30 &   0.06  &  2.18 &  6.82 & 9.76&16.55& 0.33 \\ 
NGC 5707  & 2& 13.30 &0.00738&0.14 & 0.79 & 1.77 &   0.11  &  0.70 &  3.71 &13.08&16.43& 0.18 \\ 
NGC 5719  & 2& 13.26 &0.00581&0.42 & 0.66 & 1.61 &   0.09  &  0.91 &  1.45 &12.12&14.93& 0.67 \\ 
NGC 5838  &-3& 11.92 &0.00455&0.19 & 0.70 & 2.66 &   0.05  &  0.94 &  2.78 &10.66&17.00& 0.79 \\ 
NGC 5854  &-1& 12.71 &0.00579&0.16 & 0.61 & 3.57 &   0.11  &  0.73 &  1.91 & 9.19&15.56& 0.19 \\ 
NGC 5866  &-1& 10.74 &0.00224&0.25 & 0.81 & 4.00 &   0.05  &  0.82 &  2.14 & 9.83&14.83& 0.44 \\ 
NGC 5879  & 4& 12.22 &0.00258&0.37 & 0.69 & 2.19 &   0.05  &  0.20 &  0.71 &11.56&14.81& 0.03 \\ 
NGC 5908  & 3& 12.71 &0.00579&0.24 & 0.85 & 4.47 &   0.14  &  2.14 &  5.17 & 9.21&14.73& 0.08 \\ 
NGC 5965  & 3& 12.60 &0.01138&0.39 & 0.68 & 3.94 &   0.13  &  3.20 &  6.54 & 9.56&16.14& 0.19 \\ 
NGC 5987  & 3& 12.72 &0.01004&0.34 & 0.63 & 3.53 &   0.08  &  1.97 &  5.00 & 9.13&16.06& 0.37 \\
NGC 6010  & 0& 13.60 &0.00630&0.28 & 0.74 & 2.19 &   0.25  &  0.49 &  2.19 &10.60&15.41& 0.19 \\ 
NGC 6368  & 3& 13.10 &0.00922&0.01 & 0.55 & 4.15 &   0.20  &  2.88 &  3.33 & 9.83&15.95& 0.15 \\ 
NGC 6504  & 2& 13.50 &0.01597&0.31 & 0.86 & 3.20 &   0.07  &  2.95 &  7.29 &10.19&15.98& 0.25 \\ 
NGC 6757  & 0& 13.88 &0.00710&0.10 & 0.56 & 2.68 &   0.20  &  0.39 &  2.38 &10.31&16.27& 0.13 \\ 
NGC 7311  & 2& 13.36 &0.01512&0.13 & 0.65 & 2.65 &   0.10  &  1.53 &  6.35 &10.25&16.18& 0.28 \\ 
NGC 7332  &-2& 12.02 &0.00391&0.31 & 0.76 & 4.18 &   0.09  &  0.62 &  2.22 & 6.89&15.52& 0.26 \\ 
NGC 7457  &-3& 12.09 &0.00271&0.35 & 0.54 & 4.80 &   0.10  &  1.42 &  2.38 & 8.65&16.87& 0.26 \\ 
NGC 7537  & 4& 13.86 &0.00892&0.35 & 0.65 & 1.43 &   0.16  &  0.43 &  1.79 &14.23&15.41& 0.03 \\ 
NGC 7711  &-2& 13.07 &0.01353&0.30 & 0.79 & 3.06 &   0.19  &  2.42 & 14.33 &10.21&17.89& 0.34 \\ 
\enddata
\labtab{fitvalues}
\tablecomments{The columns are $T$: Morphological type index, $m$:
apparent magnitude, $z$: redshift, $e_b$: bulge ellipticity, $e_d$:
disk ellipticity, $n$: bulge shape parameter, $\Delta n$: error in
$n$, $r_e$: bulge effective radius, $r_d$: disk scale length, $\mu_b(0)$: 
unconvolved bulge central surface brightness, $\mu_d(0)$: unconvolved 
disk central surface brightness, $B/D$: bulge-to-disk luminosity ratio.}
\tablenotetext{a}{obtained from RC3.}
\tablenotetext{b}{obtained from NED.}
\end{deluxetable}

\begin{figure}
\epsscale{1.0}
\plotone{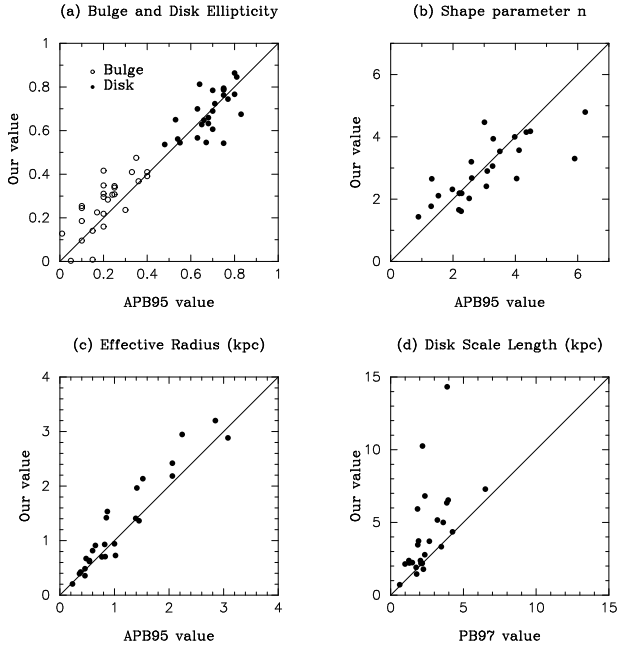}
\caption{Comparison between extracted parameters from our
decomposition program with those reported in the literature. (a) Bulge and disk
ellipticity, (b) shape parameter $n$, (c) bulge effective radius and (d) disk scale
length. Our results are compared with the results of APB95 in panels
(a),(b) and (c) and with Peletier \& Balcells (1997) in panel (d).}
\labfig{andourcomp}
\end{figure}
\newpage

\begin{figure}
\epsscale{1.0}
\plotone{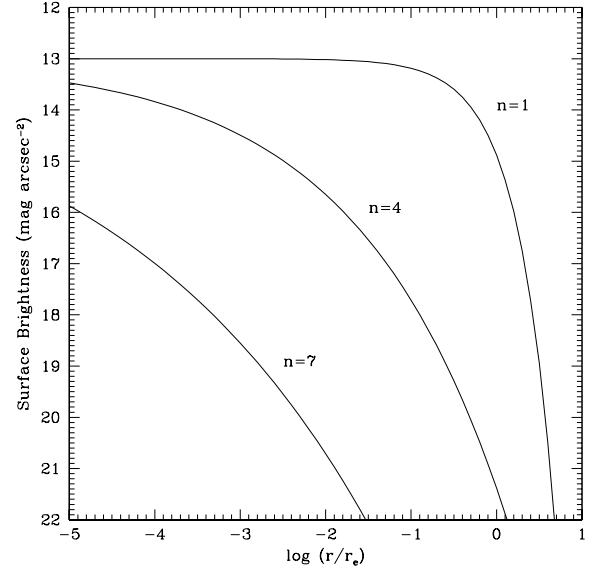}
\caption{Bulge surface brightness profiles for $n=1,4,7$. The profiles are in
arbitrary units and have been normalized to central bulge intensity $I_b(0) = 1$.}
\labfig{nprofiles}
\end{figure}
\newpage

\begin{figure}
\plotone{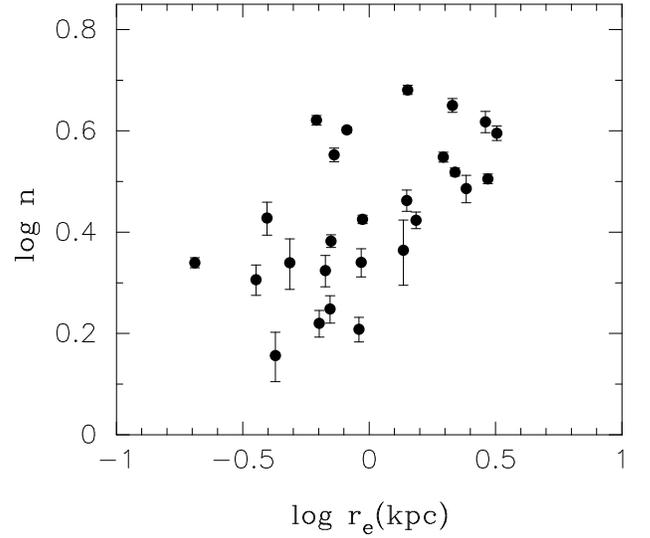}
\caption{The bulge parameter $n$  versus the bulge effective radius  in kpc. The linear correlation coefficient is 0.61 at a
significance level of 99.98 percent.}
\labfig{nrecor}
\end{figure}
\newpage

\begin{figure}
\plotone{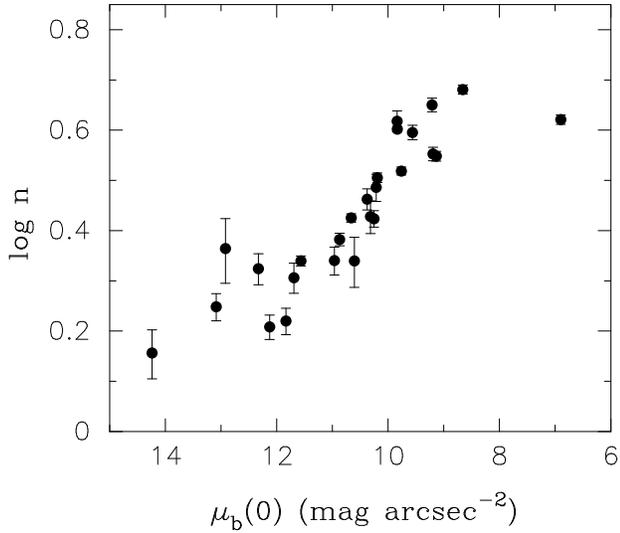}
\caption{$n$ as a function of the unconvolved bulge central
surface brightness. The linear correlation coefficient is -0.88 at a
significance level of over 99.99 percent.}
\labfig{ni0cor}
\end{figure}

\begin{figure} 
\plotone{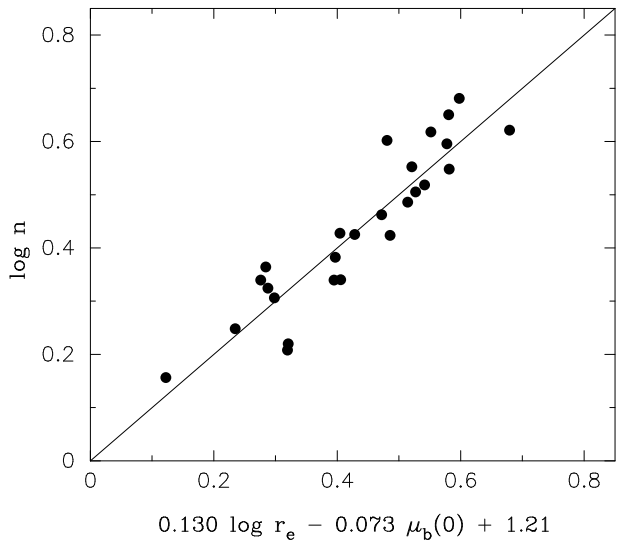}
\caption{An empirical relation 
between $n$ and the two other parameters of the bulge, $r_e$ and $\mu_b(0)$. The
solid line is the best fit line to the points. The scatter about the
line is 0.058 dex in $\log n$.}
\labfig{bicor}
\end{figure}

\begin{figure} 
\plotone{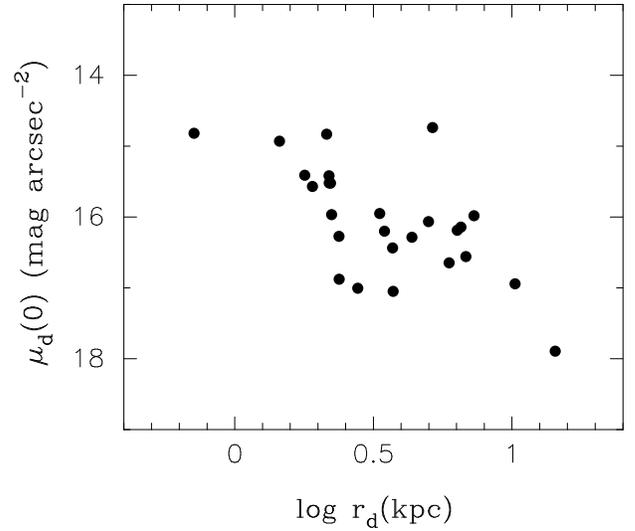}
\caption{Unconvolved disk central surface brightness $\mu_d(0)$ as a function of  disk scale
length. An anti-correlation is seen with a linear correlation coefficient
of 0.645 with a significance of 99.96 percent.}
\labfig{isrscor}
\end{figure}

\begin{figure} 
\plotone{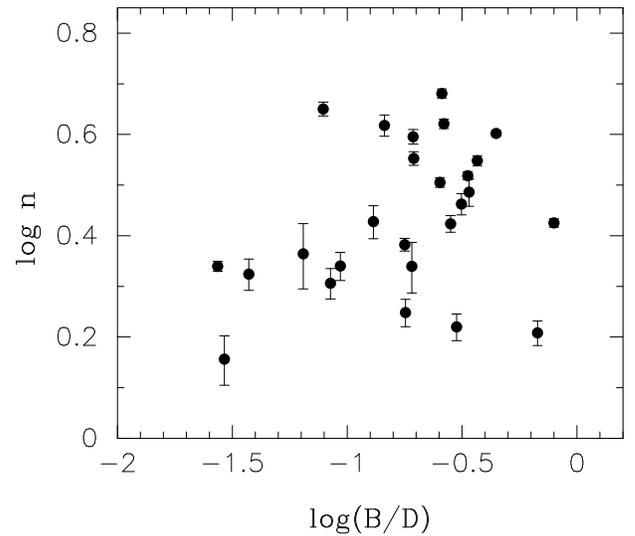}
\caption{ $n$ versus bulge-to-disk ratio. The linear correlation
coefficient is 0.30 with a significance of 86.5 percent.}
\labfig{nbdcor}
\end{figure}

\begin{figure}
\plotone{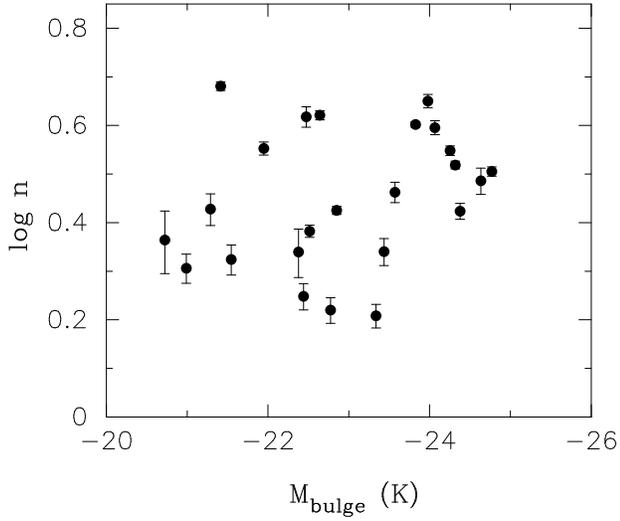}
\caption{The best-fitting  $n$ versus bulge absolute magnitude in the
K-band.  The linear correlation coefficient is -0.40 at a
significance level of 95.9 percent.}
\labfig{nabmagcor}
\end{figure}

\begin{figure} 
\plotone{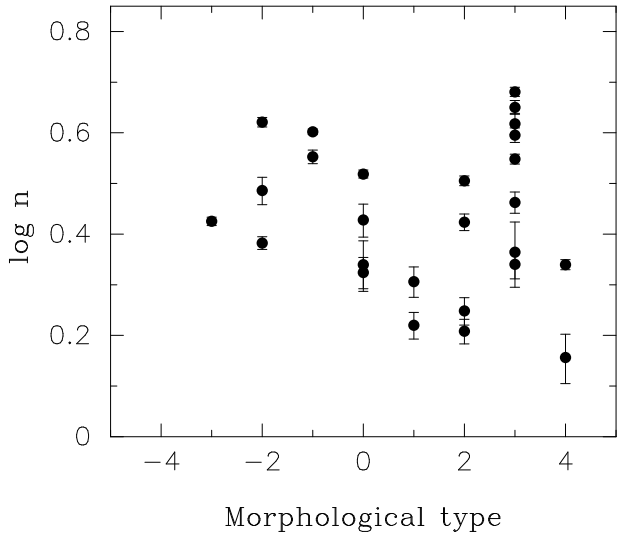}
\caption{Shape parameter $n$ versus morphological type.}
\labfig{nmorcor}
\end{figure}

\begin{figure} 
\plotone{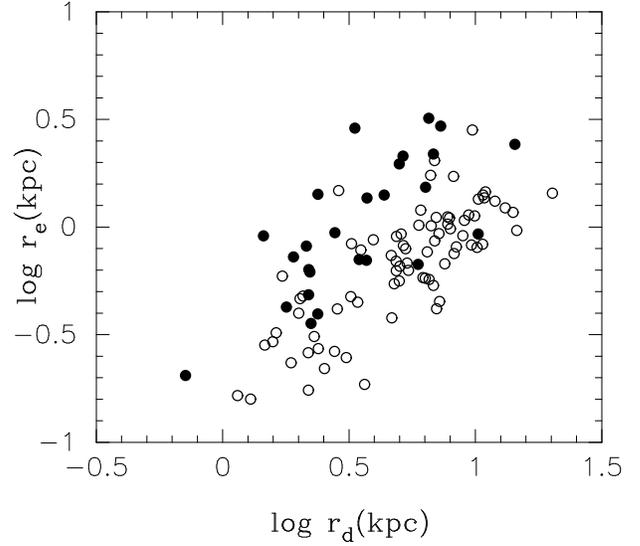}
\caption{The dependence of bulge effective radius on disk scale length. The filled
circles represent our sample, while the open circles are the $K$ band
data from
de Jong (1996).}
\labfig{rersdejong}
\end{figure}

\begin{figure} 
\plotone{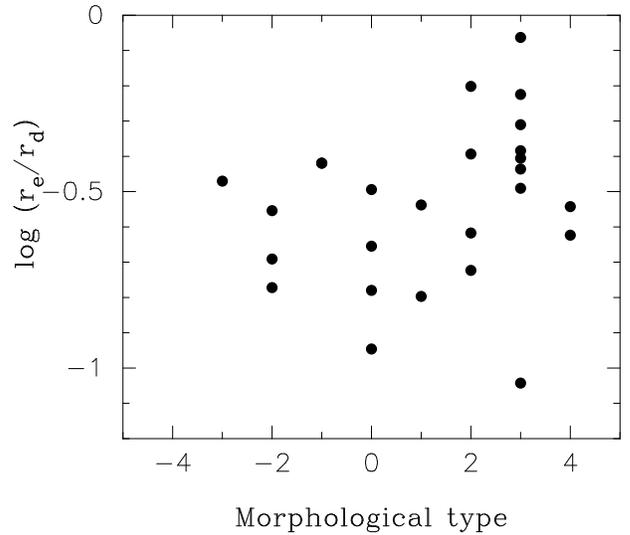}
\caption{The variation of $r_e/r_d$ ratio with
morphological type.}
\labfig{rebyrdmorcor}
\end{figure}

\begin{figure}
\plotone{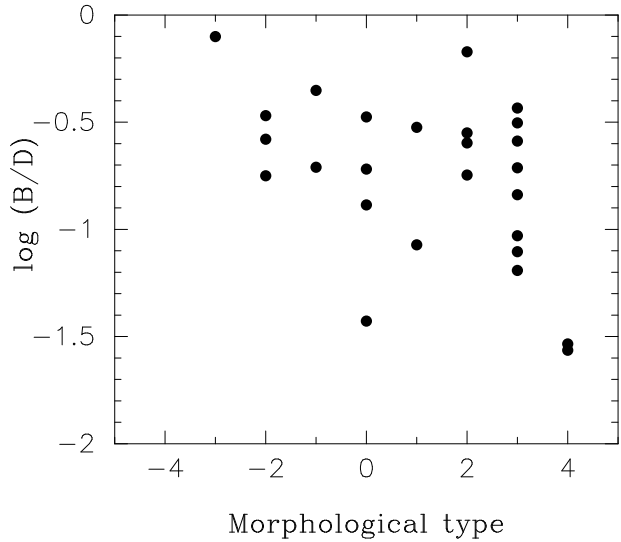} 
\caption{The variation of bulge-to-disk luminosity ratio with
morphological type. An anti-correlation characteristic of the Hubble
sequence is seen.}
\labfig{bdmorcor}
\end{figure}

\begin{figure}
\plotone{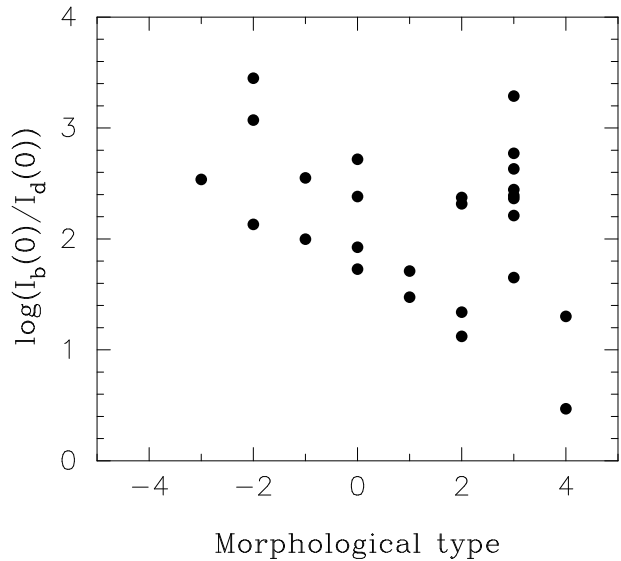} 
\caption{The dependence of bulge-to-disk central intensity ratio on
morphological type.}
\labfig{i0byidmorcor}
\end{figure}

\begin{figure} 
\plotone{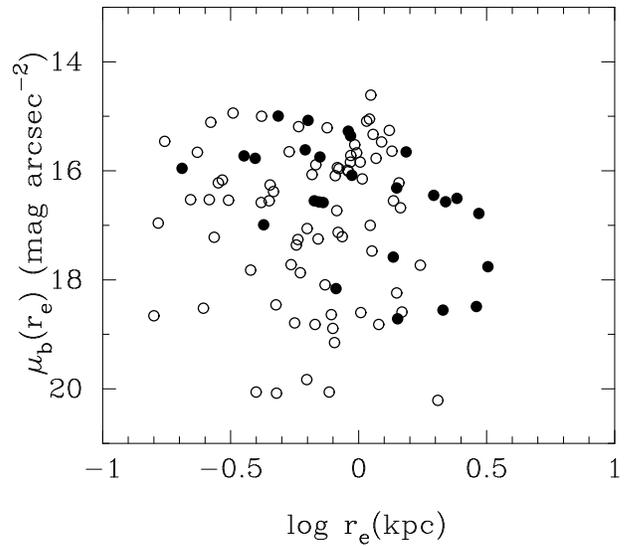}
\caption{The dependence of the surface brightness at the effective
radius $\mu_b(r_e)$  on $r_e$. The
filled circles represent our sample while the open circles are the
$K$ band data from de Jong (1996).}
\labfig{sbredejong}
\end{figure}
\end{document}